# Delay and Delay Variation Constrained Algorithm based on VNS Algorithm for RP Management in Mobile IPv6

Youssef Baddi[1] and Mohamed Dafir Ech-Cherif El Kettani[2]

[1] Information Security Research Team – ISeRT,
ENSIAS - Mohammed V-Souissi University
Rabat, Morocco
[1] *baddi@ensias.ma*

[2] Information Security Research Team – ISeRT,
ENSIAS - Mohammed V-Souissi University
Rabat, Morocco
*Dafir@ensias.ma*

**Abstract**
On account the progress of network multimedia technology, more and more real-time multimedia applications arrive with the need to transmit information using multicast communication. These applications are more important with arrival of mobile IPv6 protocol with mobile receivers and sources. These applications require a multicast routing protocol which has packets arriving at the multicast receptors within a specified QoS guaranteed and a quick recovery mechanism. In multicasting with mobile IPv6, the mobility of a receivers and senders may lead to serious problems. When the receiver or sender moves, the quality of full multicast tree may degrade so that multicast datagrams cannot be forwarded efficiently. D2V-RPM (delay and delay variation RP Manager) problem consist of choosing an optimal multicast router in the network as the root of the shared multicast tree (ST) within a specified delay and delay variation bound for all multicast session, and recovering this RP if it's off optimal or failed. This NP hard problem needs to be solved through a heuristic algorithm. In this paper, we propose a new RP Management algorithm based on Variable Neighborhood Search algorithm, based on a systematic neighborhood changing. D2V-VNS-RPM algorithm selects and recovers the RP router by considering tree cost, delay and delay variation. Simulation results show that good performance is achieved.

## 1. Introduction

Some important new emerging applications (videoconferencing) require simultaneous communication between groups of computers with quality of service (QoS) requirement. Therefore, Deering [1] proposed a technique called IP multicast routing, which entrusts the task of data duplication to the network: applications can send one copy of each packet and address it to the group of involved computers; the network takes care of message duplication to the receivers of the group. Multicast IP is a bandwidth conserving technology that reduces traffic in the network, and by the many, the bandwidth consumption also.

Multicast communication is based on a multicast tree for data routing; multicast routing protocols are built using two kinds of multicast trees: source based trees and shared trees. With source based trees, a separate tree is built for each source. With a single shared tree, one tree is built for the entire group and shared among all senders; shared trees have a significant advantage in terms of routing resources more than source-based trees in that only one routing table entry is needed for the group [2].

Shared trees require the selection of a central router called "Rendezvous point" RP in PIM-SM protocol [3] (and "core" in CBT protocol [4]). Finding out an optimal position of this router is a very well knew problem called core selection problem. Core selection directly impacts on the structure of the tree and the performance of multicast routing scheme.

Existing multicast protocols implicitly assume that group members are topologically stationary. However, the arrival of Mobile IPv6 [5] networks introduce group members and sources mobility; the group members (receivers and senders) are mobile and may move regularly from one IP network to another. Consequently, the multicast routing protocols in mobile IPv6 environment deals in same time with dynamic group memberships (member joining or leaving multicast group) and dynamic member locations change.

With the proliferation of existing multimedia group applications needs mobile group communication, the construction of Shared multicast tree satisfying the quality of service (QoS) requirements in all multicast session is becoming a problem of the prime importance, this requirements get more complicated when performing multicasting IP in mobile networks. Current implementations of PIM-SM [3] decide on the core router selection administratively [6], based only on priority and IP address of each candidate core. This leads to high cost, high delay, and high congestion.

Delay and Delay Variation core selection DDVCS correspond to the problem of finding a multicast tree with a core router satisfying the Delay and Delay Variation Bound quality of service from any source to any destination of multicast group. This problem first proposed by G. N. Rouskas and I. Baldine [7] but with a single source in the multicast tree, is an NP complete problem [2, 8, 9], which needs to be solved with a heuristic algorithm. In this paper, we introduce a novel Delay and Delay Variation core selection algorithm D2V-VNS-RPM that can improve the delay and delay variation constraint in a multi source shared tree.

This paper is organized as follows. In the next section, we describe the core selection problem. Section 3 is devoted to the description of a mathematical modeling of core selection problem. Section 4 presents the state of research of the core selection problem in the literature. Section 5 describes the proposed D2V-VNS-RPM algorithm for the RP management problem. Simulation results are reported in Section 6. Finally, Section 7 provides concluding remarks.

## 2. Background and Motivations

The main role of a multicast routing protocol consist of managing multicast groups and routing multicast messages through an optimal multicast tree in order to reach all group nodes, which facilitates the operation of multicast packet duplication. Constructing a unique multicast tree covering all multicast groups members (receivers and sources) at the same time is known by the minimum Steiner tree problem (MST) [10]; this problem is NP complete [2, 8, 9], and seeks to find a low-cost tree by minimizing cost and transmission delay. Because of the difficulties in obtaining SMT, especially in larger graphs, it is often deemed acceptable to use other optimal trees to replace SMTs through a heuristic algorithm. Multicast routing protocols are classified in two categories, as mentioned earlier (SBT and ST) [11].

Source based tree SBT or Shortest Path Tree SPT is composed of the shortest paths between the source as root and each receivers of multicast group. The main motivations behind using a source based tree SBT are the simplicity of building in a distributed manner using only the unicast routing information, and optimization of transmission delay between source and each receiver [8].

The main drawbacks of SBT are: additional costs for maintaining SPT trees, and the number of statements to be stored in the nodes complexity is $O(S * G)$ ($S$ is the number of sources and $G$ is the number of groups) [25]. The shortest path tree SPT is used by several multicast routing protocols such as DVMRP [12], MOSPF [13], and PIM-DM [14].

Source-based trees in general are mostly suitable for small scale, local-area applications. The main motivation for their use is delay optimization during multicast forwarding. They are not adapted to sparse mode situation because of the additional overhead of tree maintenance; also the scalability of source-based protocols tends to degradation in terms of network resource consumption [8].

Shared trees are more appropriate when there are multiple sources in the multicast group. Under this approach, Shared trees separate the concept of source from that of the tree root. One node in the network is chosen as the center, and the sources forward messages to the center. Like SBT tree, a shortest path multicast tree is constructed rooted at the selected RP, offering better flexibility and extensibility. And only routers on the tree need to maintain information related to group members. It gives good performance in terms of the quantity of state information to be stored in the routers and the entire cost of routing tree [12, 13, 14].

Joining and leaving a group member is achieves explicitly in a hop-to-hop way along the shortest path from the local router to core router resulting in less control overhead, efficient management of multicast path in changing group memberships, scalability and performance [1, 15]. Several multicast routing protocols in the literature use Shared tree: Protocol Independent Multicasting-Sparse mode PIM-SM [3] (and Core-Based Tree (CBT) [4]). Current implementation of PIM-SM [3] (and CBT [4] protocol) divides the tree construction problem into two sub-problems, the first is center selection problem and the second is routing problem. PIM-SM [3] (and CBT [4]) uses for center selection a special router called Bootstrap router (BSR) defined in [16], which notifies a set of candidate cores. Every node uses a Hash function to map to one core according to the address of the group; this hash function is based on router priority and his IP address. Both of these parameters do not guarantee the selection of an optimal core with any delay and delay variation guarantees. This leads to high cost, high delay, and high congestion. This problem first proposed by G. N. Rouskas and I. Baldine [17], is an NP complete problem [7, 17, 18, 19], which needs to be solved through a heuristic algorithm.

Multicast tree Construction is the first step to start a multicast session; existing multicast protocols implicitly assume that group members are topologically stationary. However, the arrival of Mobile IPv6 [5] networks generates other scenarios; the group members (receivers and senders) are mobile and may move regularly from one IP network to another. Consequently, the multicast in mobile environment deals with not only dynamic group memberships (member joining or leaving multicast group), but also dynamic member locations.

The impact of member's and sources mobility is important for multicast routing protocols that use a pre-constructed multicast tress to realize the multicast distribution. These protocols may be lead to produce a higher multicast tree cost, higher number of data packets as well as larger delay for the receivers. Member's mobility may also impact the multicast tree distribution, because every time a member's location changes the multicast tree may need to be reconstructed, resulting in large overhead. Otherwise, it may result in incorrect multicast path.

Two basic mechanisms have been proposed with Mobile IPv6 in RFC3775 [5] to ensure a multicast service and to handle multicast communications with mobile members (sources and receivers). Which are the remote subscription (RS) and the home subscription (HS) called also bi-directional tunneling. Rendezvous RP location is influenced by many events and mechanisms in mobile IPv6 environment, our motivation in this work is to overcame all of these events and mechanism to manage efficiently the Rendezvous point router in PIM-SM [3] protocol, we cite same of these events and mechanisms:

- Membership dynamics: this problem is relative to multicast IP and shared tree mechanism based on explicitly dynamic join and leave. This event can quickly degrade Rendezvous point RP quality and consequently multicast tree multicast. Since group membership can be dynamic, the Rendezvous Point manager mechanism must be able to follow current membership during all session's lifetime. Following is needed both to start forwarding data to new group members and to stop the wasteful transmission of packets to members that have left the group.
- Network dynamics: PIM-SM multicast routing protocol is independent of the unicast routing protocol, it implement its own restoration mechanism. During the lifetime of a multicast session, if any node or link supporting the multicast session fails, Rendezvous point RP quality will be degraded. This requires mechanisms to quickly recover the multicast tree by selecting other optimal Rendezvous Point RP.
- Sender handover: Many solutions are proposed to make multicast sources handover transparent to multicast tree, because a multicast session is associate a multicast address with a multicast group address, if any source change point o attachment will fail RPF check. Solutions proposed to solve RPF check failure don't overcome Rendezvous point quality degrades and also multicast tree quality degrades.
- Receiver handover: When establishing a multicast tree, existing multicast protocols implicitly assume that the group members are topologically stationary. However, in Mobile IPv6 environment, the members are mobile and may move from one IP subnet to another. Rendezvous Point RP is statically selected prior to multicast tree construction, frequent group members handover can quickly lead to a situation in which these essential multicast routers are o optimal. This situation further induces the non-optimality of the multicast routing tree. Many solution are proposed to make group members handover transparent [5, 20, 21], but it don't take quality of selected Rendezvous Point router and recovery mechanism to reselect Rendezvous Point RP.

In this paper, we propose a new RP manager Algorithm D2V-VNS-RPMbased on a "Variable Neighborhood Search". VNS algorithm has already been applied successfully to resolve a wide variety of NP-hard problems [22, 23, 24, 25, 26, 27] to select a global optimal solution using several neighborhoods structures systematically, but not yet in RP manager problem with QoS guarantee. D2V-VNS-RPM can simultaneously minimize the delay, delay variation and cost of the multicast tree. It attempts to find the best RP using a fitness function.

## 3. LITERATURE REVIEW

There are several proposals, algorithms and mechanisms for core selection problem in the literature. A variety of these algorithms are compared in [8]. Among proposed selection algorithms, we find the Random Selection, in which, the center is chosen randomly among the network. It is comparable to selecting the first source or the initiator of the multicast group, as proposed in PIM-SM [3] and CBT [4] protocols.

Some Proposed algorithms select Core on the basis of basic heuristics and do not consider QoS constraints. This kind of Core selection can provide every member of the group with a cost function guarantee to the RP iteratively selected, with this set of algorithms is hard to guarantee

QoS for all group members. From this algorithm set we mention OCBT proposed by Shields and Garcia-Luna-Aceves [28], Topology-Based Algorithm [9], which selects a single core closest to topology center by using the domain topology and sub-graph constructed from the multicast group. In order to reduce the search area used by the Topology-Based Algorithm, and to select a distributed cores for all multicast groups in the network domain and close to the group members, [9] proposed group-based algorithm. Tournament-based algorithm proposed by Shukla, Boyer, and Klinke [15] executes a Distributed tournament between nodes to determine a center; Tabu Search algorithm for RP selection (TRPSA) [29] is a distributed core selection algorithm, based on dynamic meta-heuristic Tabu Search TS algorithms proposed first by Glover [30] to solve combinatorial optimization problems in PIM-SM protocol [3]. We cite also our algorithms VNS-RP [31], VND-CS [32] and GRAS-RP [33] based on VNS [34], VND [35] and GRAS [36] heuristics successively.

There are also many well-known approaches to select core router satisfying delay and delay-variation constraints.

Delay Variation Multicast Algorithm (DVMA) was proposed by G. N Rouskas and I. Baldine [7] to resolve the Delay and Delay Variation Bounded Multicasting Network (DVBMN) problem. DVMA tries to find a sub-network given a source and a set of destinations that satisfies the QoS (Quality of Service) requirements on the maximum delay from the source to any of the destinations and on the maximum inter-destination delay variance: it starts with a source-based tree spanning some and not always all multicast members satisfying the delay constraint only. Then the algorithm searches through the candidate paths satisfying the delay and delay-variation constraints from a non-tree member node to any of the tree nodes. DVMA is most classed in source-based tree then shared tree, and it assumes that the complete topology is available at each node. The computer simulation shows that the performance of DVMA is good in terms of multicast delay-variation. However, it shows a high complexity ( $O(kldn^4)$ ) where k and l are the number of paths satisfying the delay bound between any two nodes; $|D| = d$ and $|N| = n$ represents number of multicast receptors node and total number of nodes in the topology network respectively.

Delay and Delay Variation Constraint Algorithm (DDVCA) was proposed by Sheu and Chen [17] based on the Core Based Tree (CBT) [4]: the main objective of DDVCA is to find as much as possible core router spanning a multicast tree with a smaller multicast delay variation under the multicast end-to-end delay constraint.

To do that, DDVCA first calculates the delay of the least delay path from the destination nodes to all the nodes. The node that has the minimum delay-variation is selected as the core node. In comparison with the DVMA, DDVCA Algorithm shows a significant lower complexity i.e. $O(dn^2)$ where m is the number of destination nodes and n is the total number of nodes in the computer network.

KIM et.al [37] has proposed another efficient core selection algorithm based also on CBT like DDVCA [17] to build a core based multicast tree under delay and delay-variation bound. First, AKBC finds a set of candidate core nodes that have the same associated multicast delay-variation for each destination node. Then, it selects a final core node from this set of candidate core nodes that has the minimum potential delay-variation. AKBC algorithm investigates candidate nodes to select the better node with the same complexity as DDVCA i.e $O(dn^2)$.

All these algorithms (DDVCA [17], DVMA [7] and AKBC [37]) are only applied in the symmetric network environment that has no direction. To overcome this limitation, Ahn, Kim and Choo [19] proposed AKC (Ahn Kim Choo) to build a multicast tree with low delay-variation in a realistic network environment that has two-way directions. This algorithm works eficiently in the asymmetric network with the same complexity as DDVCA i.e $O(dn^2)$.

The last core selection algorithm, proposed by Sahoo and. al [38], is based on dynamic meta-heuristic Tabu Search TS algorithms, proposed first by Glover [31], to solve combinatorial optimization problems. Tabu Search algorithm for RP selection (TRPSA) [38] is a distributed core selection algorithm to find a local solution after a certain finite number of iterations by using memory structures that describe the visited solutions. The basic idea of the TRPSA [7] algorithm is to mark the best local solution obtained in order to prevent the research process to return back to the same solution in subsequent iterations using a data structure to store the solutions already visited, this structure is called tabu list. However, the method requires a better definition of stopping criterion and effective management of the tabu list, since the choice of stopping criterion and tabu list size is critical and influences the performance of the algorithm. According to [38], TRPSA has $O(|E| + (|S| + |D|)|N|^2)$ complexity.

However, these algorithms [17, 31, 25, 1] select the best core node out of a set of candidate core nodes that have the same associated delay-variation. Therefore, these algorithms are restricted only to selecting the best core node, which may not generate an optimal delay-variation-based multicast tree in many cases. Also TRSPA [38] doesn't overcome this limitation because it just selects a local optimal node which may not generate an optimal delay and delay variation-based multicast tree in all topology networks.

## 4. MATHEMATICAL MODELING

A computer network is modeled as a simple directed and connected graph $G = (N, E)$, where $N$ is finite set of nodes and $E$ is the set of edges (or links) connecting the nodes. Let $|N|$ be the number of network nodes and $|E|$ the number of network links. An edge $e \epsilon E$ connecting two adjacent nodes $u \epsilon N$ and $v \epsilon N$ will be denoted by $e(u,v)$, the fact that the graph is directional, implies the existence of a link $e(v,u)$ between v and u. Each edge is associated with two positive real value: a cost function $C(e) = C(e(u,v))$ represents link utilization (may be either monetary cost or any measure of resource utilization), and a delay function $D(e) = D(e(u,v))$ represents the delay that the packet experiences through passing that link including switching, queuing, transmission and propagation delays. We associate for each path $P(v_0, v_n) = (e(v_0,v_1), e(v_1,v_2), \cdots, e(v_{0n-1}, v_n))$ in the network two metrics:

$$C\big(P(v_0, v_n)\big) = \sum_{0}^{n-1} C(e(v_i, v_{i+1}))$$

and

$$D\big(P(v_0, v_n)\big) = \sum_{0}^{i+1} C(e(v_i, v_{i+1}))$$

A multicast tree $T_M(S, RP, D)$ is a sub-graph of $G$ spanning the set of sources node $S \subset N$ and the set of destination nodes $D \subset N$ with a selected Rendezvous Point RP. Let $|S|$ be the number of multicast destination nodes and $|D|$ is the number of multicast destination nodes.

In Protocols using Core-based tree as PIM-SM [3] protocol, all sources node needs to transmit the multicast information to selected core via unicast routing, then its well be forwarded to all receptors in the shared tree, to model the existence of these two parts separated by core, we use both cost function and delay following:

$$C\big(T_M(S, RP, D)\big) = \sum_{0}^{n-1} C\big(e(s_i, RP)\big) + \sum_{d \in D} C(e(RP, d))$$

And

$$D\big(T_M(S, RP, D)\big) = \sum_{0}^{n-1} D\big(e(s_i, RP)\big) + \sum_{d \in D} D(e(RP, d))$$

We introduce also a Delay Variation in formula 3 function defined as the difference between the Maximum and minimum of end-to-end delays (formulas 1 and 2) along the multicast tree from the source to all destination nodes and is calculated as follows:

$$MaxDelay = \max(D\big(T_M(S, RP, D)\big)) \quad (1)$$
$$MinDelay = \min(D\big(T_M(S, RP, D)\big)) \quad (2)$$
$$DelayVariation = MaxDelay - MinDelay \quad (3)$$

Core selection problem tries to find an optimal node as a Rendezvous Point RP in the network with an optimal function $Opt\_F$ by minimizing in the first time the cost function $C\big(T_M(S, RP, D)\big)$ and in the second a Delay and delay variation bound as follows:

$$Opt\_F(RP, T\_M) = \begin{cases} \min(C\big(T_M(S, RP, D)\big)) \\ Delay \leq \alpha \\ DelayVariation \leq \beta \end{cases} \quad (4)$$

## 5. D2V-VNS-RPM ALGORITHM

### 5.1 Basic Variable Neighborhood Search VNS algorithm

Contrary to all others kind of meta-heuristics based on local search methods, Mladenović and Hansen [39] proposed a recent meta-heuristics Variable Neighborhood Search VNS Algorithm based on the simple idea of a systematic neighborhood changing arbitrarily, which vary in size, but usually with increasing cardinality, within a local search algorithm (Hill Climbing, Simulated annealing, tabu ...).

VNS has been applied successfully to a wide variety of NP hard problems to select a global optimal solution such as the travelling salesman problem [23], Job Shop Scheduling Problems [24], the clustering problem [26], Arc routing problems [23], and nurse rostering [25].

The use of more than one neighborhood provides a very effective method that allows escaping from a local optimum. In fact, it is often the case that the current solution, which is a local optimum in one neighborhood, is no longer a local optimum in a different neighborhood; therefore, it can be further improved using a simple descent approach.

As defined by Mladenović and Hansen [39] and presented in fig. 1, in the VNS paradigm, a finite set of neighborhoods structures $N_K$ ($K = 1, \cdots, K_{max}$) and an initial solution $S$ are generated, starting from this initial solution, a so-called shaking step is performed by randomly selecting a solution $S'$ from the first neighborhood, This is followed by applying an iterative improvement local search algorithm to get a $S''$ Solution. If this solution $S''$ improves the weight function defined in formula (4) one starts with the first neighborhood of this new solution $S \leftarrow S''$; otherwise one proceeds with the next neighborhood. This procedure is repeated as long as a neighborhood structure allows such iteration.

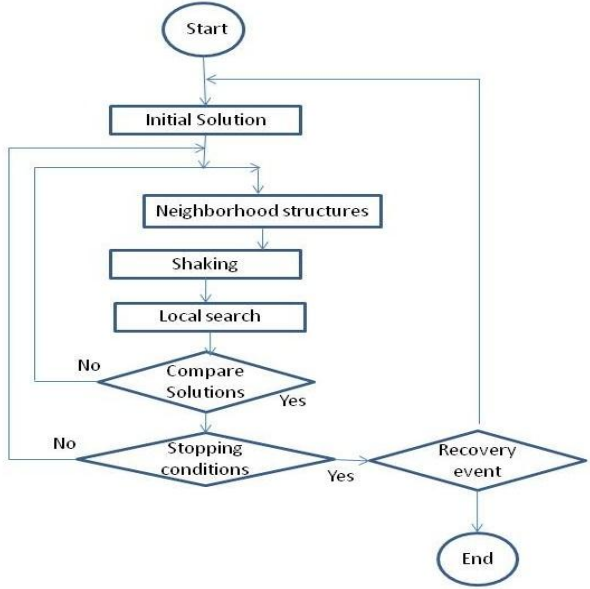

Fig. 1: D2V-VNS-RPM algorithm execution

### 5.2 A variable neighborhood descent for core selection problem

The main motivation behind the use of the VNS search algorithm to solve core selection problem is the use of several neighborhoods to explore different neighborhood structures systematically. Our goal is to break away from a local minima, this use is based on three facts:
1. If node $N_1$ is a local minimum for one neighborhood structure $N_K$ is not necessary so with another one $N_{K'}$.
2. A global minimum solution S is a local minimum for all possible neighborhood structures.
3. For the core selection problem local minima for all neighborhood structures is relatively close and localized in the same place.

In this section we provide a detailed description of the Variable Neighborhood Search algorithm for RP selection Problem with delay and delay variation guarantee D2V-VNS-RPM, and his three process phases: the initialization process, stopping conditions phase, the shaking step and recovery phase.

### 5.3 Initial solution

The first step of variable neighborhood search is to define an initial solution. Many methods can be used to generate this solution; the simplest is to select randomly one node in the network as initial solution. There are other methods that try to reduce the selection area and generate an initial solution from an ordered set of multicast group members.

In this paper we use the selected node as a RP by the bootstrap Protocol as an initial solution.

### 5.4 Neighborhood structures

D2V-VNS-RPM uses neighbor nodes concept to generate neighborhood structures: a node $u$ is neighbor of another node $v$ if an edge $e(u,v)$ between $u$ and $v$ exists. We propose to compute a neighborhood structure $N_j$ through the following formula (with $\text{neighbor}(S)$ a set of neighbor nodes of $S$):

$$N_j(S) = \begin{cases} neighbor(S), & if\ j = 1 \\ \bigcup_{x\ \epsilon\ N_{j-1}(S)} N(x), & else \end{cases}$$

### 5.5 Shaking

From an initial solution *S*, the shaking function varies and explore a new parts of the search space in random manner. It chose other solution $S'$ from the $K^{th}$ neighborhood structure $N_K(s)$. For this reason, Neighborhood structures are ranked in such a way that VNS algorithm explores increasingly further away from $S'$. After exploring search space thoroughly with a local search algorithm, the best local solution $S''\ is\ selected$, $S''$ is compared with *S*. If $S''$ is better than *S*, it replaces *S* ($S \leftarrow S''$) and the algorithm starts all over again with $k \leftarrow 1$. Otherwise, $k$ is incremented and algorithm continues from shaking phase with next neighborhood structure. This function of shaking is terminated after all neighborhood structures are exhausted. The result of each Shake $S'$ is used as the starting point for the next Local Search.

### 5.6 Local search phase

In recent years, several local search algorithms have been proposed, they proceed from an initial solution *S* generated randomly from a neighborhood structure and trays to find an optimum local solution $S''$ by a sequence of local changes, which improve each time the value of the objective function. The D2V-VNS-RPM is independent of de local search algorithm used; it can work with hill climbing [40], adaptive multi-start [41], simulated annealing [42], Tabu search (TS) [30], GRASP [36]...

### 5.7 Stop conditions

An iteration of the algorithm D2V-VNS-RPM is composed principally of a shaking function, running a local search algorithm and a test of movement. In case of small problem instances, where the best solution is usually found very quickly, the stopping condition with a limit on the maximum number of iterations is sufficient. Therefore, a second stopping condition has been added for large-scale problems. This criterion is the maximum number of iterations after obtaining solutions with the same _F .

5.8 Recovery phase

The main challenge to integrate multicast IP in mobile IPv6 networks is the rapidly changing environment. Hence, it's difficult to maintain an optimal Rendezvous point RP for all multicast group members and sources during the multicast session. In this section we present a relocation mechanism in which RP could be relocated periodically and after a event change to preserve an optimal shared multicast tree.

Based on a fixed timestamp value, D2V-VNS-RPM initiate
RP selection algorithm periodically, it uses also signaling messages to maintain multicast tree by initiate RP selection after a network events change.

D2V-VNS-RPM uses different signaling messages of PIM-SM multicast protocol such as hello, join/prune, RP candidature messages to detect any change in unicast and multicast states i.e. Link Failure, Router Failure.... It use also Mobile IPv6 message such as mobile ipv6 binding to detect receivers and sources mobility.

5.9 VNS-RP algorithm and pseudo code

In this section, a step by step VNS-RP-search-based algorithm for RP selection problem is presented, also we present a pseudo code in algorithm 1.

**Step 1:** Declaration of provided information according to the network graph G.
**Step 2:** Set maximum iteration number of VNS, maximum number of iteration of stable solution, maximum iteration number of local search method...
**Step 3:** Select initial solution S.
**Step 4:** Choose the $K_{max}$ scalar, select the set of neighborhood structures $N_K$, for $K = 1, \cdots, K_{max}$, that will be used in the search; choose a stopping condition.
**Step 5:** Shaking phase: Take at random a solution $S$ from $N_K(S)$.
**Step 6:** Local search phase: execute a local search algorithm, such as tabu search [30], GRASP [36] ..., to produce a local optimal solution S.
**Step 7:** Check if objective function value of solution $S''$ is less than objective function value of solution S, then move to $S''$ solution and continue the search with $N_K$ ($K \leftarrow 1$) from step 4; otherwise, set $K \leftarrow K + 1$ and also continue the search from step 4;
**Step 8:** Output the best solution core selected.
**Step 9:** Waiting for a recovery event

| Algorithm 1: VNS-RP Pseudo code |
|---|
| **Input**: i = 0 |
| **Input**: totalIt = 0 |
| /*number of iteration of stable solution */ |
| **Input**: maxItWithoutImprovement |
| **Input**: initialSolution |
| **Input**: Solutionsheking |
| **Input**: Solutionlocal |
| 1  **while** events **do** |
| 2   **while** i < maxItWithoutImprovement && totalIt < max it **do** |
| 3    k ← 0 ; |
| 4    **while** current it < max it && k < k_max **do** |
| 5     **if** totalIt > max it **then** |
| 6      break ; |
| 7     **end** |
| 8     getN_k(s) ; |
| 9     Solutionsheking ← getRandomN_k(s) ; |
| 10    Solutionlocal←localSearchMethod(Solutionsheking); |
| 11    **if** opt Function(Solutio local) > opt Functions(s) **then** |
| 12     k ← k + 1 ; |
| 13    **else** |
| 14     s ← Solution local; |
| 15     k ← 0 ; |
| 16    **end** |
| 17    totalIt ← totalIt + 1; |
| 18    current it ← current it + 1; |
| 19   **end** |
| 20   **if** lastSCost > opt Function(s)) **then** |
| 21    i ← 0; |
| 22   **else** |
| 23    i ← i + 1; |
| 24   **end** |
| 25  **end** |
| 26  **return** s ; |
| 27 **end** |

## 6. SIMULATION RESULTS

In this section, we use simulation results to demonstrate the effectiveness of the proposed algorithm described above. To study the performance of our selection algorithm D2V-VNS-RPM, we implement it in a simulation environment; we use the network simulator NS2 [43] with mobiwan [44] module to implement Mobile IPv6 [5]. The random graph generator GT-ITM [45] is used to generate a random different 100 networks, and we adopt Waxman [46] as the graph model.

Our simulation studies were performed with a 100 runs. The values of $\alpha = 0.2$ and $\beta = 0.2$ were used to generate networks with an average degree between 3 and 4 in the mathematical model of Waxman.

To demonstrate the performance of this algorithm (D2VVNS-RPM), we compare it with the following algorithms, including random (R), Tabu RP selection (TRPS) [38], DDVCA [17] and AKC [19].

The main objective of our algorithm is to reduce delay and delay variation during all multicast session; therefore, we start the simulation results by comparing these two metrics.

For this we use a topology network with multicast group member's size is 10 % of the overall network nodes. Delay is defined as the time spent to forward multicast data from the source node to the furthest receiver node in the multicast group. Figure 2 shows the simulation results of multicast delay versus the number of nodes in the topology network. Simulation results show that D2V-VNS-RPM is the best among all the algorithms on average delay, with TRPS and AKC following it, and DDVCA and Random are the worst.

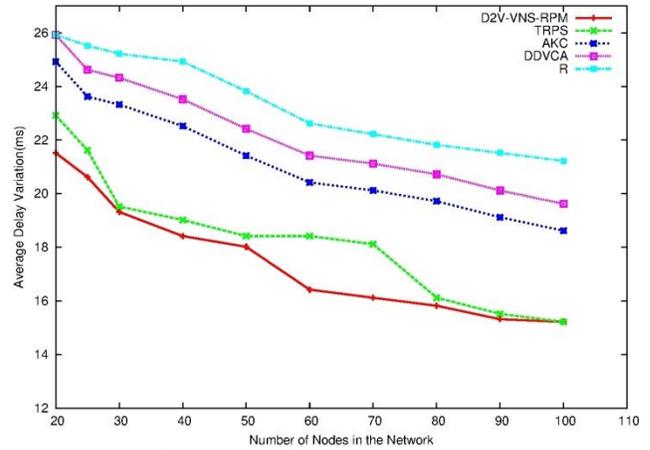
Fig. 3: Comparison of Delay Variation vs. network size

Based on the cost function in the formula 5, Figure 4 presents a comparison study of multicast tree Cost generated by each algorithm, the performance of Random selection is the worst, followed by AKC , DDVCA and, D2V-VNS-RPM shows better performances, and it has the minimal cost, this cost is also reserved during all multicast session, because the multicast tree is update when a recovery event is signaled.

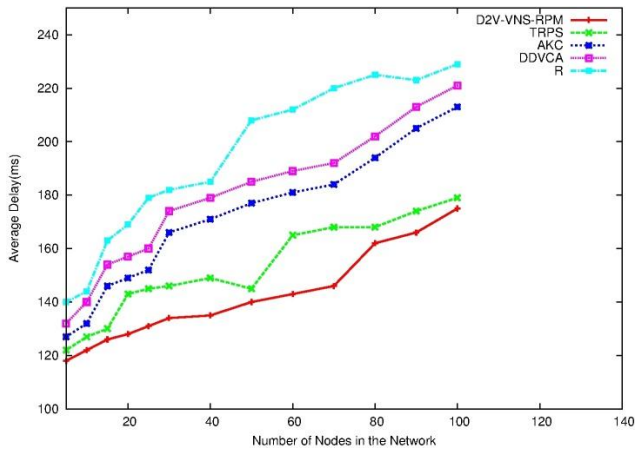
Fig. 2: Comparison of Delay vs. network size

Delay Variation is the difference between the first time of the reception of a multicast packet by a receiver of the multicast group and the last reception of the same multicast packet by another receiver of the multicast group. In Figure 3 the Delay Variation is plotted as a function of the number of nodes in the network topology, it shows that D2V-VNS-RPM decrease more the delay variation to transmit multicast packet to all multicast group, this reduction is caused by the selection of an optimal Rendezvous Point RP using the formula 3, followed by others algorithms.

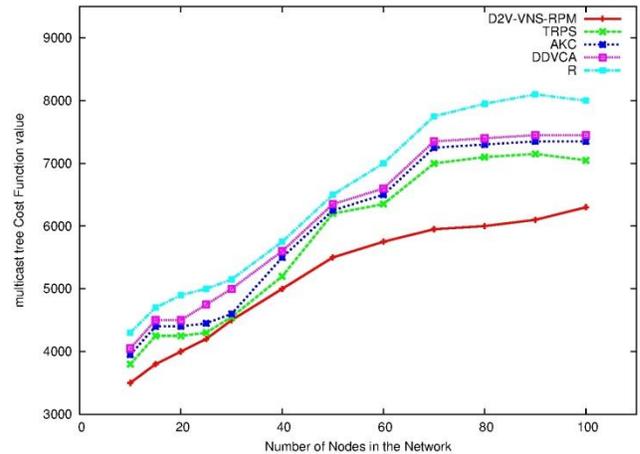
Fig. 4: Comparison of multicast tree Cost vs. network size

Packets lost is very important metrics in multicast session with a multimedia application, it's also a very important metrics in mobile environment with a mobile multicast members and sources. We have studied the packet losses caused by multicast sources and members handoff and by all multicast events. Figure 5 shows the average of Packets lost with the number of nodes in the network topology, it shows that D2V-VNS-RPM reduce more the Packets lost followed by AKB, DDVCA and TRPS. This reduction is get because D2V-VNS-RPM us a recovery events to maintain multicast tree and by selecting an optimal Rendezvous Point to avoid congestion problem.

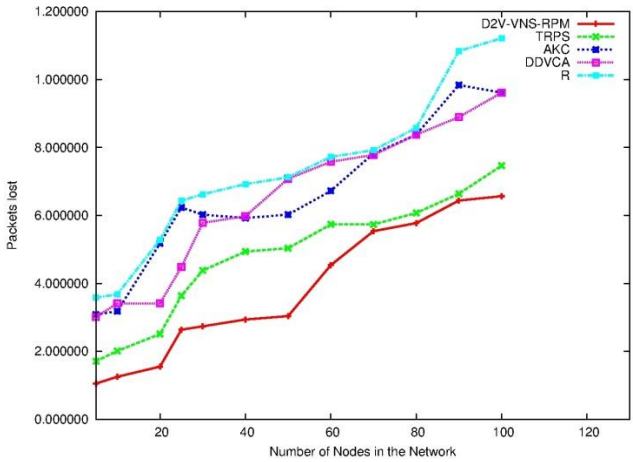
Fig. 5: Comparison of Average Packet Lost vs. network size

Handoff Latency is defined for a mobile receiver as the time that elapses between the last time T1 a data packet received via the old route before handover and the time T2 when the first data packet is received along the new route after a handover. Handoff latencies affect the service quality of mobile communication and multicast diffusion, especially, with real-time applications. Figure 6 shows the variation of Handoff Latency as a function of the nodes number in the network topology, it shows that D2V-VNS-RPM reduce the Handoff latency time to transmit multicast packet to all multicast members, followed by AKB, DDVCA and TRPS. This reduction is due to quick tree recovery with an optimal Rendezvous Point.

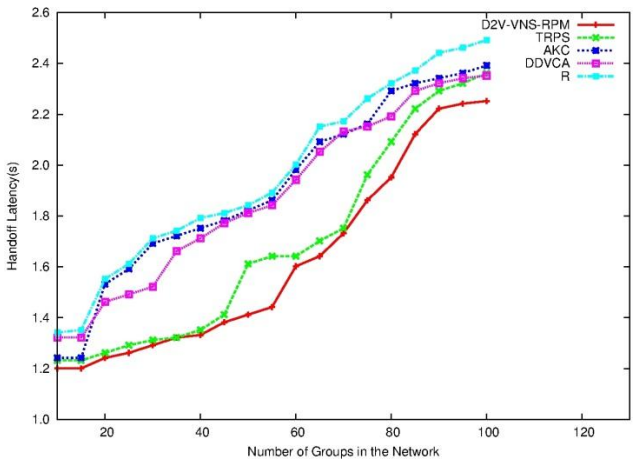
Fig. 6: Comparison of Handover Latency vs. network size

Through handover from one IP subnet to another, the mobile source node needs to send multicast packets continuously while moving. However, the handover is unpredictable and there is no forwarding mechanism for multicast traffic addressed to mobile members, consequently a handover event generate a 100% of packets lost. This sources and members handovers cause also an off optimal multicast tree. Figure 7 shows Impact of mobility in Packets lost, it shows that D2V-VNS-RPM reduce more the Packets lost generated in all sources and members handover; it shows also that AKB, DDVCA and TRPS generate more Packets lost caused by time required to select an optimal Rendezvous Point RP and so the multicast tree rebuild.

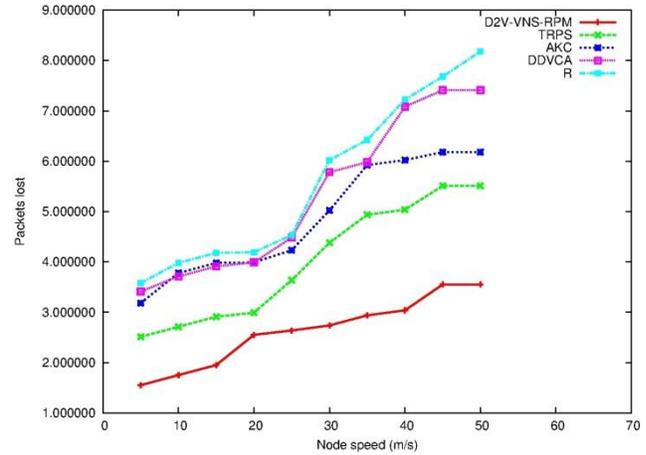
Fig. 7: Comparison of Average Packet Lost vs. mobility Speed

The main objective to using multicast routing scheme is to reduce multicast packet in the network, our proposed algorithm try to select an optimal Rendezvous Point RP, consequently, an optimal multicast tree with less Throughput Consumed. Figure 8 shows the variation of Throughput Consumed with the number of nodes in the network topology, it shows that D2V-VNS-RPM is the best scheme to integrate multicast IP with Mobile IPv6; it reduces the Throughput Consumed and the number of packet in the network, followed by AKB, DDVCA and TRPS. This is because other algorithms need more information to select the Rendezvous Point RP router.

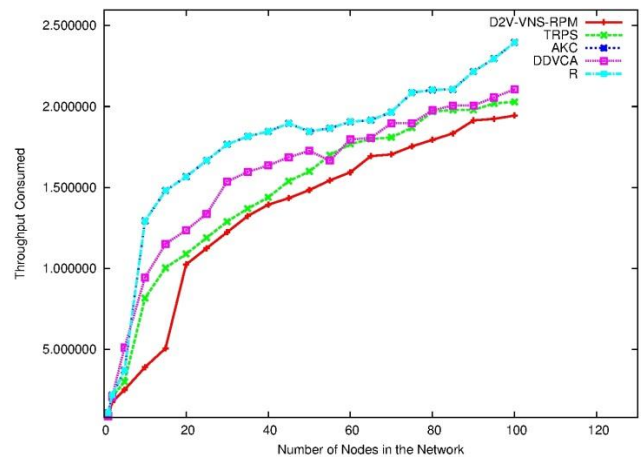
Fig. 8: Comparison of Throughput Consumed vs. network size

## 4. CONCLUSION

In this paper we have investigated the problem of finding and selecting an optimal Rendezvous Point RP router during all multicast session in a Mobile IPv6 environment. RP Selection problem in PIM-SM [3] multicast routing protocol affects directly the structure of the tree and the performance of the routing scheme of multicast consequently. Current algorithms decide on core router administratively, which leads to high cost, high delay, and high congestion. To solve these problems, D2V-VNS-RPM is proposed based on VNS heuristic algorithm. To present our proposition we started with a brief overview of multicast routing protocols and two types of multicast trees SBT and ST. We reviewed and analyzed the cost, delay and delay variation functions. We reviewed the RP selection algorithms studied in literature. Simulation results show that our algorithm presents good performances in multicast cost, delay, and delay variation

## References

[1] S. E. Deering and D. R. Cheriton. Multicast routing in datagram internetworks and extended lans. ACM Transactions on Computer Systems, 8:85-110, 1990.

[2] D. Zappala, A. Fabbri, and V. M. Lo. An evaluation of shared multicast trees with multiple cores. Telecommunication Systems, 19(3-4):461-479, 2002.

[3] B. Fenner, M. Handley, H. Holbrook, and I. Kouvelas. Protocol Independent Multicast - Sparse Mode (PIM-SM): Protocol Specication (Revised). RFC 4601 (Proposed Standard), August 2006. Updated by RFC 5059.

[4] A. Ballardie. Core based trees (cbt version 2) multicast routing - protocol specification -. RFC 2189, 1997.

[5] D. Johnson, C. Perkins, and J. Arkko. RFC 3775: Mobility Support in IPv6, June 2004.

[6] M. Ramalho. Intra- and inter-domain multicast routing protocols: A survey and taxonomy. IEEE Communications Surveys and Tutorials, 3(1):2-25, 2000.

[7] G. N. Rouskas and I. Baldine. Multicast routing with end-to-end delay and delay variation constraints. Technical report, Raleigh, NC, USA, 1995.

[8] A. Karaman and H. Hassanein. Core-selection algorithms in multicast routing - comparative and complexity analysis. Comput. Commun., 29(8):998-1014, 2006.

[9] K. L. Calvert, E. W. Zegura, and M. J. Donahoo. Core selection methods for multicast routing. Pages 638{642, 1995.

[10] K. Mehlhorn. A faster approximation algorithm for the steiner problem in graphs. Inf. Process. Lett., 27:125-128, March 1988.

[11] L. Wei and D. Estrin. The trade-offs of multicast trees and algorithms. 1994.

[12] D. Waitzman, C. Partridge, and S. E. Deering. RFC 1075: Distance vector multicast routing protocol, Nov. 1988. Status: EXPERIMENTAL.

[13] J. Moy. MOSPF: Analysis and Experience. RFC 1585 (Informational), Internet Engineering Task Force IETF, March 1994.

[14] D. Farinacci, T. Li, S. Hanks, D. Meyer, and P. Traina. Protocol independent multicast – dense mode (pim-dm): Protocol specification (revised). RFC 3973, 2005.

[15] S. B. Shukla, E. B. Boyer, and J. E. Klinker. Multicast Tree Construction in Network Topologies with Asymmetric Link Loads. PhD thesis, 1994.

[16] N. Bhaskar, A. Gall, J. Lingard, and S. Venaas. Bootstrap Router (BSR) Mechanism for Protocol Independent Multicast (PIM). RFC 5059 (Proposed Standard), January 2008.

[17] P.-R. Sheu and S.-T. Chen. A fast and efficient heuristic algorithm for the delay- and delay variation bound multicast tree problem. In Proceedings of the The 15th International Conference on Information Networking, ICOIN '01, pages 611{, Washington, DC, USA, 2001. IEEE Computer Society.

[18] Y. Ahn, M. Kim, Y.-C. Bang, and H. Choo. On algorithm for the delay- and delay variation-bounded multicast trees based on estimation. In L. T. Yang, O. F. Rana, B. D. Martino, and J. Dongarra, editors, Proceedings of the First international conference on High Performance Computing and Communications, volume 3726 of HPCC'05, pages 277-282, Berlin, Heidelberg, 2005. Springer-Verlag.

[19] S. Ahn, M. Kim, and H. Choo. Efficient algorithm for reducing delay variation on delay-bounded multicast trees in heterogeneous networks. In WCNC 2008, IEEE Wireless Communications and Networking Conference, March 31 2008 - April 3 2008, Las Vegas, Nevada, USA, Conference Proceedings, pages 2741-2746. IEEE, 2008.

[20] C. Jelger and T. NoËl. Supporting mobile ssm sources for ipv6. 2:1693{1697, Nov 2002.

[21] A. O'Neill. Mobility Management and IP Multicast. Internet Draft - work in progress (expired) 01, IETF, July 2002.

[22] M. Tagmouti, M. Gendreau, and J.-Y. Potvin. A variable neighborhood descent heuristic for arc routing problems with time-dependent service costs. Comput. Ind. Eng., 59:954{963, November 2010.

[23] F. Carrabs, G. Laporte, and J. Cordeau. Variable neighborhood search for the pickup and delivery traveling salesman problem with LIFO loading. Publication (Centre for Research on Transportation (Montréal, Québec))). Centre for Research on Transportation = Centre de recherche sur les transports (C.R.T.), 2005.

[24] G. N. Rouskas and I. Baldine. Multicast routing with end-to-end delay and delay variation constraints. Technical report, Raleigh, NC, USA, 1995.

[25] E. Burke, P. De Causmaecker, S. Petrovic, and G. V. Berghe. Variable neighborhood search for nurse rostering problems, pages 153{172. Kluwer Academic Publishers, Norwell, MA, USA, 2004.

[26] P. Hansen and N. Mladenović. Variable neighborhood search: Principles and applications. European Journal of Operational Research, 130(3):449{467, May 2001.

[27] P. Hansen, N. Mladenović, and J. A. Moreno-Pérez. Variable neighbourhood search: methods and applications. Annals OR, 175(1):367{407, 2010.

[28] C. Shields and J. J. Garcia-Luna-Aceves. The ordered core based tree protocol. In Proceedings of the INFOCOM '97. Sixteenth Annual Joint Conference of the IEEE Computer and Communications Societies. Driving the Information Revolution, INFOCOM '97, pages 884{, Washington, DC, USA, 1997. IEEE Computer Society.

[29] W. Hua, M. Xiangxu, Z. Min, L. Yanlong, et al. Tabu search algorithm for RP selection in PIM-SM multicast routing. Comput. Commun., 33(1):35-42, January 2010.

[30] F. Glover. Tabu search - part II. INFORMS Journal on Computing, 2(1):4{32, 1990.

[31] Y. Baddi and M. D. El Kettani, "VND-CS: a variable neighborhood descent algorithm for core selection problem in multicast routing protocol," in Fourth International Conference on Networked Digital Technologies 2012 (NDT 2012), Dubai, UAE, Apr. 2012.

[32] Y. Baddi and M. D. El Kettani, "VNS-RP algorithm for rp selection in multicast routing protocol pim-sm," in The 3rd International Conference on Multimedia Computing and Systems (ICMCS'12), Tangier, Morocco.

[33] Y. Baddi and M. D. El Kettani, "GRAS-RP: greedy randomized adaptive search algorithm for rp selection in pim-sm multicast routing," in Seventh International Conference On Intelligent Systems: Theories And Applications (SITA'12), Mohammedia, Morocco.

[34] Hansen, P., Mladenović, N., Variable neighborhood search: Methods and recent applications. In: In Proceedings of MIC99. pp. 275–280 (1999)

[35] P. Hansen and N. Mladenović, "Variable neighborhood search: Principles and applications," European Journal of Operational Research, vol. 130, pp. 449–467, 2001.

[36] T. A. Feo and M. G. Resende, "Greedy randomized adaptive search procedures," Journal of Global Optimization, vol. 6, pp. 109–133, 1995.

[37] M. Kim, Y.-C. Bang, H.-J. Lim, and H. Choo. On efficient core selection for reducing multicast delay variation under delay constraints. IEICE Transactions, 89-B(9):2385{2393, 2006.

[38] S. P. Sahoo, M. R. Kabat, and A. K. Sahoo. Tabu search algorithm for core selection in multicast routing. In Proceedings of the 2011 International Conference on Communication Systems and Network Technologies, CSNT '11, pages 17{21, Washington, DC, USA, 2011. IEEE Computer Society.

[39] P. Hansen, N. Mladenovic, and L. C. D. Gerad. Variable neighborhood search: Methods and recent applications. In In Proceedings of MIC'99, pages 275{280, 1999.

[40] W. Buntine. A guide to the literature on learning probabilistic networks from data. IEEE Trans. On Knowl. and Data Eng., 8(2):195-210, Apr. 1996.

[41] R. Marti. Multi-start methods. In F. Glover and G. Kochenberger, editors, Handbook of Metaheuristics, volume 57 of International Series in Operations Research and Management Science, pages 355-368. Springer New York, 2003.

[42] R. Bouckaert. Bayesian Belief Networks: from Construction to Inference. PhD thesis, Utrecht, Netherlands, 1995.

[43] T. Issariyakul and E. Hossain. Introduction to Network Simulator NS2. Springer Publishing Company, Incorporated, 1 edition, 2008.

[44] T. Ernst. MobiWan: A ns-2.1b6 Simulation Platform for Mobile IPv6 in Wide Area Networks. Motorola Labs Paris, Gif-sur-Yvette, France, May 2001.

[45] H. Tangmunarunkit, R. Govindan, S. Jamin, S. Shenker, and W. Willinger. Network topologies, power laws, and hierarchy. Technical report, 2001.

[46] B. M. Waxman. Routing of multipoint connections. Selected Areas in Communications, IEEE Journal on, 6(9):1617-1622, Aug. 2002.

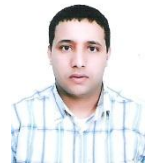
**Youssef Baddi** received MASTER Thesis degree in computer science from ENSIAS School, University Mohammed V Souissi of Rabat, Morocco, in 2010. Since 2011 he is a Ph.D. Student ENSIAS School, member of Information Security Research Team.
His research interests include multicast routing protocol, IPv6, mobile IP, Integration of Multicast IP and mobility IPv6 management, with emphasis on the algorithms for Core selection in the Shared Multicast trees protocol.

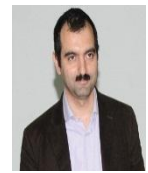
**Mohamed Dafir Ech-Chrif El Kettani** received his engineering degree in 1994 and Ph.D. degree from EMI School, University Mohammed V Agdal of Rabat in 2001. He is currently a professor at ENSIAS School, University Mohammed V Souissi of Rabat, Morocco and Information Security Research Team leader. Currently, he is also the IT Manager of University Mohammed V Souissi since 2011. His main research interests are in the area of Security management and multicast routing.